\documentclass[11pt]{article}
\usepackage{axodraw}
\usepackage{pstricks}
\usepackage{color}
\usepackage{graphicx}
\usepackage{slashed}

\newcommand{\be}{\begin{equation}}
\newcommand{\ee}{\end{equation}}
\newcommand{\ba}{\begin{array}{c}}
\newcommand{\ea}{\end{array}}
\newcommand{\bqa}{\begin{eqnarray}}
\newcommand{\eqa}{\end{eqnarray}}
\makeatletter
    
    \newcommand{\Rmnum}[1]{\expandafter\@slowromancap\romannumeral #1@}
    \makeatother

\begin{document}

\begin{center}
{\Large\bf Production of $X(3872)$ at PANDA}
\vskip 10mm
G.Y. Chen$^1$ and J.P. Ma$^2$   \\
{\small {\it $^1$ Department of Physics, Peking University,
Beijing 100871, China }} \\
{\small {\it $^2$ Institute of Theoretical Physics, Academia Sinica,
Beijing 100080, China }} \\
\end{center}
\begin{abstract}
The recently discovered $X$(3872) has many possible interpretations.
We study the production of $X$(3872) with PANDA at GSI for the antiproton-proton
collision with two possible interpretations of X(3872).
One is as a loosely-bound
molecule of $D$-mesons, while another is a 2P charmonium
state $\chi_{c1}$ (2P).
Using effective couplings we are able to give numerical predictions
for the production near the threshold and the production associated
with $\pi^0$. The produced $X$(3872) can be identified with its decay
$J/\psi \pi^+\pi^-$. We also study the possible background
near the threshold production for $X(3872) \to J/\psi \pi^+\pi^-$.
With the designed luminosity $1.5{\rm fb}^{-1}$ per year of PANDA we find that the event number
of $p\bar p \to J/\psi \pi^+\pi^-$ near the threshold is at the order of $10^6 \sim 10^8$,
where the large uncertainty comes from the total decay width of $X(3872)$. Our study
shows that at the threshold more than about $60\%$ events come from the decay of $X(3872)$
and two interpretations are distinguishable from the line-shape of the production.
With our results  we except that the PANDA experiments will shed light on
the property of $X(3872)$.
\end{abstract}

\vspace{2cm}
$X(3872)$ has been first discovered by Belle collaboration\cite{BELL} in the decays
$B\rightarrow K X(3872)\rightarrow K J/\psi \pi^+\pi^-$. Later,
its existence has been confirmed by experiments of Babar\cite{Babar}, CDF\cite{CDF}
and D0\cite{D0}. The word average mass of
X(3872) now is $m_X=(3871.2\pm 0.5)$MeV and the total width is $\Gamma_X
< 2.3$MeV at 90\% C.L.\cite{PDG}. The angular distribution analysis made by Belle
\cite{BQ} favors $J^{PC} =1^{++}$. A similar analysis by CDF
\cite{CDQ} collaboration allow $J^{PC} = 1^{++}$ and $J^{PC} =
2^{-+}$ as well. The dipion mass distribution in the decay into  $J/\psi \pi^+\pi^-$
suggests that the $\pi^+\pi^-$ may come from a $\rho$ resonance, this
is supported by the CDF analysis\cite{CDQ}.
\par
Many interpretations of $X(3872)$ exist. In \cite{Dbs} it is interpreted
as a loosely-bound molecule of $D^0 \bar{D}^{\ast
0}-c.c.$. In \cite{CM,PC} it is suggested that $X(3872)$
is the first excited state of the conventional charmonium $\chi_{c1}$, i.e.,
$\chi_{c1}(2P)$. Other possible interpretations are also possible,
like the $S$-wave threshold effect of $D^0 \bar{D}^{\ast
0}$\cite{JR}, a cusp effect\cite{DB}, a diquark anti-diquark bound state\cite{MPPR},
a hybrid charmonium state\cite{BL} and a tetraquark state\cite{tetra}, etc.
The existence of these many interpretations reflects the fact that the structure
of $X(3872)$ is still unclear. It is clear that further studies in experiment and
theory are needed.
\par
In this work we study the production of $X(3872)$ in $p\bar p$ collisions
by taking $X(3872)$ as a  loosely-bound molecule of $D^0 \bar{D}^{\ast
0}-c.c.$ or as the first excited state of the conventional charmonium $\chi_{c1}$.
Experimentally the production can be studied with PANDA detector
for the anti-proton beam facility at GSI\cite{PANDA}, where the anti-proton
is with the energy from $1 \sim 15$GeV.
In $p\bar p$ collisions $X(3872)$ can be produced near its threshold.
We assume it will be identified through its decay into $J/\psi \pi^+ \pi^-$,
then the same final state can also be produced through direct production,
which will be a background in identification of $X(3872)$.  We will make numerical predictions
for the process $p\bar p \to J/\psi \pi^+ \pi^-$ near the threshold
of $X(3872)$, where the final state is produced through the decay of $X(3872)$ or
through the direct production. We will also give numerical results for the
production associated with a $\pi^0$. Theoretical study of the $X(3872)$ production
at quark-gluon level in the energy range we consider is very difficult.
We will take the approach of effective Lagrangian in terms of hadrons.
We first discuss couplings between relevant hadrons and then
give our numerical results.
\par
If we assume the X(3872) is a pure 2P charmonium state $\chi_{c1}$
(2P), then we can estimate it decay width of  into $p\bar p$ as
following. In the decay the charm quark pair will be annihilated into
gluons first, then those gluons will be converted into the $p\bar p$ pair.
The conversion will be the same for $\chi_{c1}$ in the ground and the first
excited state. We take charm quarks as heavy quarks and use
a nonrelativistic wave function  to describe the charm quark pair
in the charmonia. In the nonrelativistic limit, the annihilation rate
of $\chi_{c1}$ into gluons will be proportional to the square of the first derivative
of the radial wave-function $R(r)$ . Therefore we have:
\begin{equation}
 \frac{\Gamma[X(3872)\rightarrow p\bar
p]}{\Gamma[\chi_{c1}\rightarrow p\bar
p]}=\frac{| R^\prime(0)|^2_{\chi_{c1}(2P)}}{|R^\prime(0)|^2_{\chi_{c1}}}.
\end{equation}
One can obtain the wave functions with some potential models.
In \cite{EC} the
numerical results for four different potentials are given. Here we use the
result with the Cornell potential\cite{EC}:
\begin{equation}
\frac{|R^\prime(0)|^2_{\chi_{c1}(2P)}}{|R^\prime(0)|^2_{\chi_{c1}}}=\frac{0.186}{0.131}=1.42.
\end{equation}
From other three models the ratio is $0.97$, $1.05$ and $1.33$, respectively. One can re-scale
our prediction for the total cross-section with the ratio from the Cornell model to
obtain the prediction with ratios from other three models. It should be noted
that in \cite{EC} the main quantum number is defined as $n_r + \ell +1$. Therefore
the $2P$ state in \cite{EC} is the $P$-wave ground state while the $3P$ state is the first
excited $P$-wave state.
Using the above results and experimental data we can determine
the effective coupling constant $g_{p\bar p X}$ which is defined as
\begin{equation}
{\mathcal L}_{p\bar p X} = g_{p\bar p X}\bar p \gamma^\mu \gamma_5 p X_\mu,
\ \ \ \ \ g_{p\bar
pX}=1.11\times10^{-3},
\end{equation}
where $X_\mu$ is the effective field for $X(3872)$.

\par
If X(3872) is a loosely-bound molecule of $D^0
\bar{D}^{\ast 0}$, the decay  width into $p\bar p$ has
been estimated by~\cite{EB2} as:
\begin{equation}
\Gamma[X(3872)\rightarrow p\bar
p]=\left (\frac{\Lambda}{m_{\pi}}\right )^2\left (\frac{E_X}{0.6{\rm MeV}}\right )^{1/2} (35 {\rm eV}),
\ \ \ \ \  E_X=M_{D^0\bar{D^{\ast 0}}}-m_X,
\end{equation}
where $\Lambda$ can be chosen as $m_{\pi}$ since low-energy
scattering of charm mesons is dominated by pion exchange and
$E_X=0.6\pm0.6$ MeV is the bounding energy of the molecular state.
We use $\Lambda=m_{\pi}$, $E_X=0.6$ MeV  and have for the effective
coupling:
\begin{equation}
 g_{p\bar pX}=7.14\times10^{-4}.
\end{equation}
\par
Having fixed the coupling with $p\bar p$ we turn to the decay $X\to J/\psi \pi^+\pi^-$.
As discussed at the beginning, it is likely that the $\pi$-pair comes
from the $\rho$-resonance. We will take the decay
as $X\to J/\psi \rho \to J/\psi \pi^+\pi^-$. Then decay amplitude
with effective couplings can be written as:
\begin{eqnarray}
\mathcal {M}[X\rightarrow J/\psi\pi^{+}\pi^{-}]&=& \mathcal
{A}_{\mu\alpha}[X\rightarrow J/\psi
\rho] \epsilon_{X}^{\alpha} \frac{-g^{\mu\nu}}{q^2-M_{\rho}^2+iM_{\rho}\Gamma_{\rho}}
\mathcal {A}_{\nu}[\rho \rightarrow \pi^{+}\pi^{-}],
\nonumber\\
 \mathcal
{A}_{\nu}[\rho\rightarrow\pi^{+}\pi^{-}] &=& \frac{1}{2}G_{\rho\pi\pi}(p_{+}-p_{-})_{\nu},
\nonumber\\
 \mathcal {A}_{\mu\alpha}[X\rightarrow
J/\psi\rho]& = & G_{X\psi\rho}\varepsilon_{\mu\nu\alpha\beta}q^{\nu}
\epsilon_{\psi}^{\ast\beta},
\end{eqnarray}
where $q$ is the four momentum of $\rho$,
$p_{+}$ and $p_{-}$ are the momentum of $\pi^{+}$,$\pi^{-}$,
respectively, and $\epsilon_{X}$,$\epsilon_{\psi}$ are the
polarization four vector of the X(3872) and the $J/\psi$.  The
coupling constant $G_{\rho\pi\pi}$ can be determined from the decay
of $\rho$ into $\pi^{+}\pi^{-}$, which is $11.99\pm0.06$.
The total decay width can be obtained as:
\begin{equation}
 \Gamma[X\rightarrow
J/\psi\pi^+\pi^-]=|G_{X\psi\rho}|^2 (226 {\rm keV}),
\end{equation}
where we have used a cutoff for the invariant mass of the $\pi$-pair, which is taken
as $m_{2\pi} > 400$MeV as in the experiment of Belle\cite{BELL}.
\par
If X(3872) is a loosely-bound state of the charm mesons,
the coupling $G_{X\psi\rho}$ can be expressed with the total width
and binding energy\cite{EB1}:
\begin{equation}
|G_{X\psi\rho}|^2\approx0.86
\left (\frac{E_{X}+\Gamma_X^2/(16E_X)}{0.7{\rm MeV} } \right )^{1/2},
\end{equation}
where $\Gamma_X$ is the total width of
the $X(3872)$ and the lower bound on
width to be $\Gamma_X
> 2\Gamma[D^{\ast 0}]=136\pm32$keV\cite{EB1}.
If we take $E_X=0.6$MeV and  the upper bound
to be $2.3$ MeV, then we obtain
\begin{equation}
  G_{X\psi\rho}\approx
0.893\sim 1.05.
\end{equation}
For the case that $X(3872)$ is a $2P$ charmonium state $\chi_{c1}(2P)$,
the decay width is estimated as\cite{CM}:
\begin{equation}
\Gamma[X\rightarrow J/\psi\pi^+\pi^-]=40
{\rm keV},
\end{equation}
which gives the value of the effective coupling:
\begin{equation}
G_{X\psi\rho}\approx 0.42 .
\end{equation}
\par
With the estimated coupling constants between relevant hadrons we are able to predict
the cross-section for the process $p\bar p \to J/\psi \pi^+ \pi^-$ near the threshold
of $X(3872)$, where the final state can be produced from the decay of $X(3872)$
or from direct production. The amplitude for the final state from the decay
can be written as:
\begin{eqnarray}
\mathcal {M}[p\bar p\rightarrow  X \to J/\psi\pi^{+}\pi^{-}]&= & \mathcal
{A}_{\mu}[p\bar p\rightarrow
X]\frac{g^{\mu\nu}-P^{\mu}P^{\nu}/m_X^2}{P^2-m_X^2+im_X\Gamma_X}\mathcal
{A}_{\nu\alpha}[X\rightarrow
J/\psi\rho]
\nonumber\\
    && \cdot
\frac{ig^{\alpha\beta}}{q^2-m_{\rho}^2+im_{\rho}\Gamma_{\rho}}\mathcal
{A}_{\beta}[\rho\rightarrow \pi^{+}\pi^{-}],
\nonumber\\
 \mathcal {A}_{\mu}[p\bar p\rightarrow X] &= & g_{p\bar
pX}\bar{v}_{\bar p\bar s}\gamma_{\mu}\gamma_{5}u_{p s},
\end{eqnarray}
where $P$ denotes the momentum of the $p\bar p$-pair, and $q$ is the
momentum of the $\pi$-pair. The final state can also be produced
directly from the $p\bar p$-annihilation as shown in Fig.1.,
this should be taken as a background for the production of $X(3872)$.
\par
\begin{figure}[hbt]
\begin{center}
\includegraphics[width=10cm]{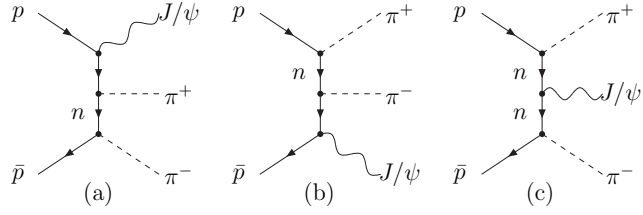}
\end{center}
\caption{ The production of $J/\psi \pi^+ \pi^-$ through the $p\bar p$-annihilation }
\label{Feynman-dg1}
\end{figure}
\par
In Fig.1 $n$ denotes the internal neutron line. Again we use effective couplings
to calculate the process.
We use $-\sqrt{2}
g_{pp\pi}\gamma_5$ for the $n p \pi$ effective vertex, $-
g_{pp\pi}\gamma_5$ for the $p p \pi$ effective vertex, $i g_{p\bar
p\psi}\gamma_{\mu}$ for the $J/\psi p\bar p$, and $i g_{n\bar
n\psi}\gamma_{\mu}$ for the $J/\psi n\bar n$, respectively.
The amplitude from Fig.1 can be expressed as
\begin{eqnarray}
 \mathcal {M}_a & = & i
2g_{pp\pi}^2g_{p\bar p\psi}\bar{v}_{\bar p\bar
s}\gamma_5\frac{1}{(\slashed{p}_{-}-\slashed{\bar
p})-m_n}\gamma_5\frac{1}{(\slashed{p}-\slashed{p}_{\psi})-m_p}\gamma_{\mu}u_{ps}\epsilon_{\psi}^{\ast\mu},
\nonumber\\
 \mathcal {M}_b & = & i 2g_{pp\pi}^2g_{p\bar p\psi}\bar{v}_{\bar p\bar
s}\gamma_{\mu}\frac{1}{(\slashed{p}_{\psi}-\slashed{\bar
p})-m_p}\gamma_5\frac{1}{(\slashed{p}-\slashed{p}_{+})-m_n}\gamma_{5}u_{ps}\epsilon_{\psi}^{\ast\mu},
\nonumber\\
\mathcal {M}_c & = & i 2g_{pp\pi}^2g_{n\bar n\psi}\bar{v}_{\bar p\bar
s}\gamma_5\frac{1}{(\slashed{p}_{-}-\slashed{\bar
p})-m_n}\gamma_{\mu}\frac{1}{(\slashed{p}-\slashed{p}_{+})-m_n}\gamma_{5}u_{ps}\epsilon_{\psi}^{\ast\mu},
\end{eqnarray}
then the total amplitude is the sum:
\begin{equation}
\mathcal {M}[p\bar p \to J/\psi\pi^{+}\pi^{-}] =
\mathcal {M}[p\bar p\rightarrow  X \to J/\psi\pi^{+}\pi^{-}]
+\mathcal {M}_a+ \mathcal {M}_b + \mathcal {M}_c.
\end{equation}
The effective coupling $g_{p p\pi}$ is $g_{p p\pi}=13.5$.
By using isospin symmetry we have $g_{p\bar p\psi}=g_{n\bar n\psi}$ and $m_n =m_p$.
The effective coupling $g_{p\bar p\psi}$ can be determined from the decay $J/\psi\to p\bar p$.
It should be noted that for the decay it is possible that another coupling, i.e.,
the Pauli's coupling can be appear\cite{BAR}. We neglect this coupling and
get $g_{p\bar p\psi}=1.62\times 10^{-3}$~\cite{BAR}.
Although the coupling constants are estimated, but their relative phase is unknown.
There are two possible cases:
Case 1: The product $g_{p\bar p\psi}g_{pp\pi}^2$ has the same sign as that
of the product $g_{p\bar pX}G_{X\psi\rho}G_{\rho\pi\pi}$. Case 2:
The two products have different sign.
The expression of the amplitude squared is too length because it is
a $2\to 3$ body process and we do not try to produce
an analytical expression for the total cross section. Instead giving
the analytical expression we simply take the amplitude squared
to perform the phase space integral numerically.
In Fig.2 and Fig.3 we plot the total cross section as functions
of the invariant mass $s$ of the $p\bar p$ pair, where we take
$\Gamma_X=2.3$ MeV and the coupling constants estimated before.
Fig.2 is for $X(3872)$ as a loosely bound state of $D$-mesons,
Fig.3 is for $X(3872)$ as $\chi_{c1}(2P)$.
\par
\begin{figure}[hbt]
\begin{center}
\includegraphics[width=8cm]{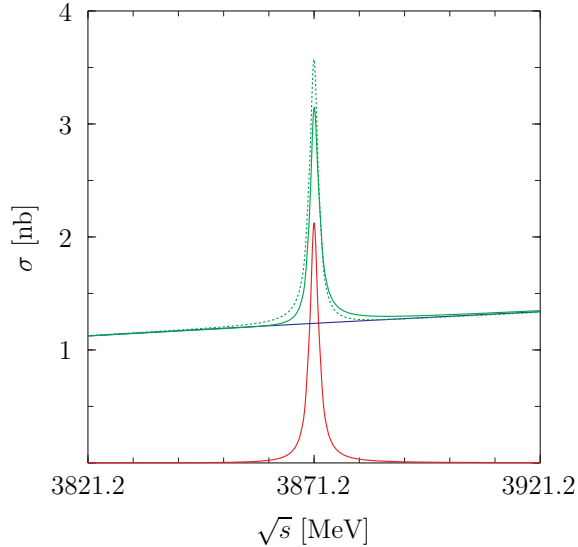}
\end{center}
\caption{The total cross-section as a function of $\sqrt{s}$
for $X(3872)$ as a loosely bound state of $D$-mesons. The lower resonance
curve is without the background, the upper curves
are with the back ground. The solid one is for Case 2, while the dashed one
is for Case 1.}
\label{Feynman-dg2}
\end{figure}
\par

\par
\begin{figure}[hbt]
\begin{center}
\includegraphics[width=8cm]{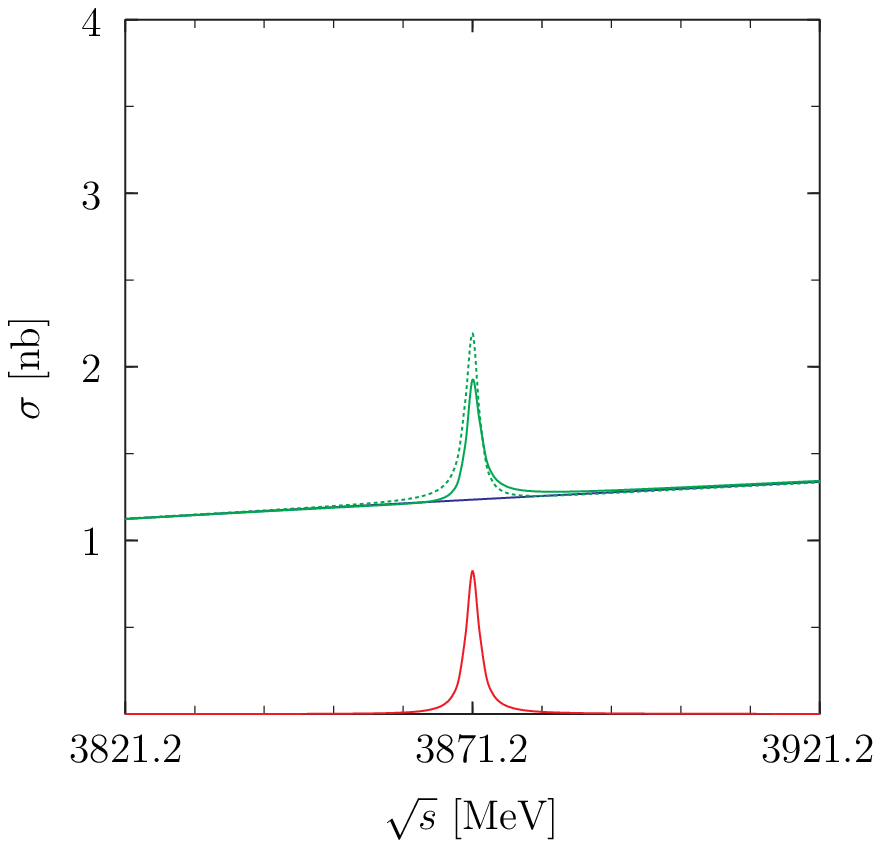}
\end{center}
\caption{The total cross-section as a function of $\sqrt{s}$
for $X(3872)$ as $\chi_{c1}(2P)$. The lower resonance
curve is without the background, the upper curves
are with the back ground. The solid one is for Case 2, while the dashed one
is for Case 1. }
\label{Feynman-dg3}
\end{figure}
\par

From these figures we clearly see that from the line shape of the cross-section
two interpretations of $X(3872)$ can be distinguished. The difference
comes from the decay into $J/\psi \pi^+ \pi^-$ with the different assignment
of $X(3872)$. We also see that the background is an significant contribution for
the production.  For the assignment with the bound state of $D$-mesons
we have the total cross-section at $\sqrt{s} =3.872$GeV
by taking $E_X=0.6$ MeV, $\Gamma_X=136 {\rm keV} \sim 2.3 {\rm MeV}$:
\begin{equation}
\sigma[p\bar p\rightarrow X \to  J/\psi\pi^{+}\pi^{-}]=3.57\sim
443{\rm nb},
\end{equation}
for the assignment with $\chi_{c1}(2P)$ we have with $\Gamma_X=136 {\rm keV} \sim 2.3
{\rm MeV}$:
\begin{equation}
 \sigma[p\bar p\rightarrow X \to  J/\psi\pi^{+}\pi^{-}]=2.19\sim
238{\rm nb} .
\end{equation}
The main uncertainties in the above come from the unknown width $\Gamma_X$.
In Fig.4 we plot the dependence of the cross section with different assignments
as a function of $\Gamma_X$. From Fig.4 we see that there is a strong dependence of the total cross section
on the total width and the cross-section with two interpretations are different.
Hence the measurement of the cross section will give a clear evidence
to indicate which interpretation is the correct one.
\par
\par
\begin{figure}[hbt]
\begin{center}
\includegraphics[width=10cm]{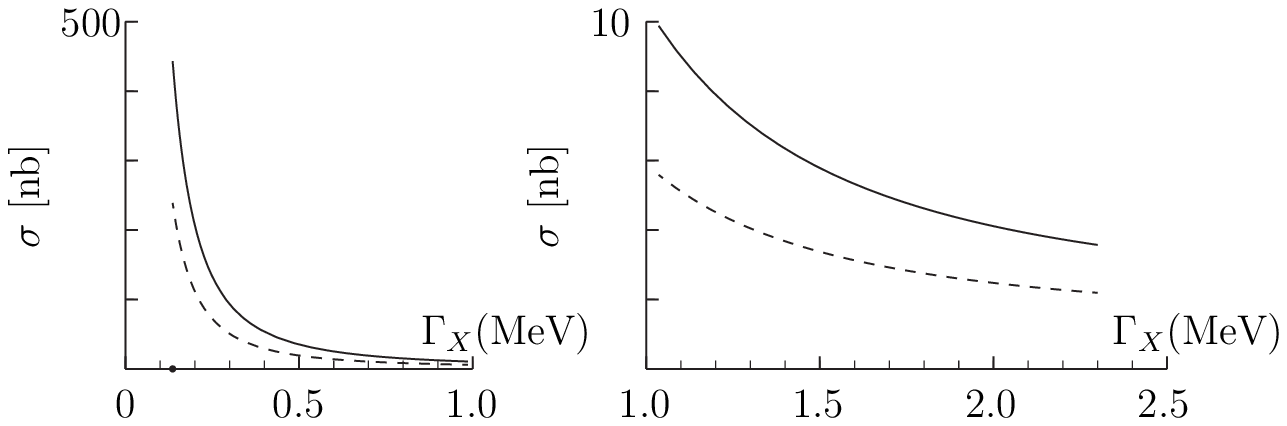}
\end{center}
\caption{The total cross-section as a function of $\Gamma_X$
at the threshold. The solid line is for the interpretation
with the $D$-meson bound state, the dashed line is for
the interpretation of $\chi_{c1}(2P)$. The curves are for Case 1.
The curves for Case 2. are similar.}
\label{Feynman-dg4}
\end{figure}

\par
The luminosity of PANDA can be up to $2\times10^{32} {\rm cm}^{-2}{\rm s} ^{-1}$ ~\cite{PANDA}. Assuming
$50\%$ overall efficiency and 6 months/year data taking,
the integrated luminosity is to be $1.5 {\rm fb}^{-1}$ per year.
With the integrated luminosity and the cross-section obtained here, one can
expect $10^6 \sim 10^8$ events per year for the production of $J/\psi \pi^+\pi^-$
near the threshold. With the large number of events one can study $X(3872)$ in more detail.
\par
With the estimated couplings we can also study the production of $X(3872)$
associated with $\pi^0$, i.e., the production away from the resonance region.
There are two diagrams for the process given in Fig.5. The calculation of the total cross section
is straightforward. The analytical expression for the differential cross-section
can be found in \cite{BAR}.
We only give our numerical result here. With
the same parameters we obtain the total cross section as a function of $\sqrt{s}$
up to $5$GeV given in Fig.6. From Fig.6 we find that the total cross section
of $p\bar p \to X \pi^0$ is at order of $\sim 100$pb.
With the designed luminosity and by considering
the branching ratio of decays of $X$ it is likely that such a process can not be observed.

\par
\begin{figure}[hbt]
\begin{center}
\includegraphics[width=8cm]{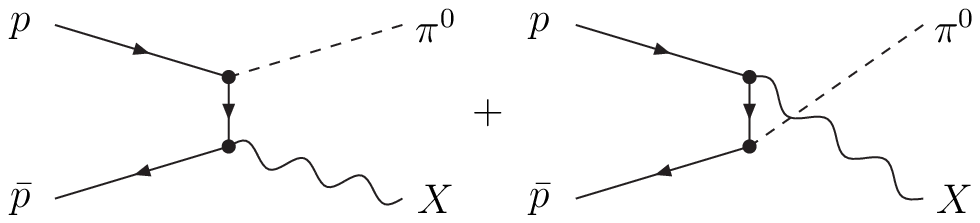}
\end{center}
\caption{The diagrams for $p\bar p \to X \pi^0$.}
\label{Feynman-dg3}
\end{figure}
\par
\par
\begin{figure}[hbt]
\begin{center}
\includegraphics[width=8cm]{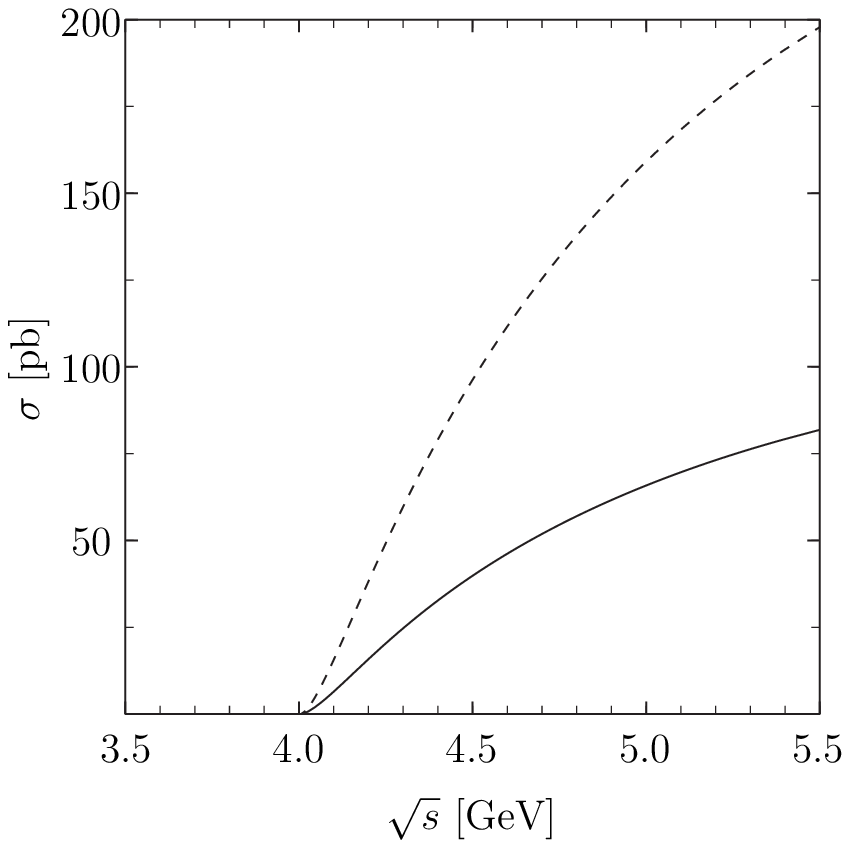}
\end{center}
\caption{The $s$-dependence of the total cross section of $p\bar p \to X \pi^0$.}
\label{Feynman-dg3}
\end{figure}
\par

To summarize: In this work we have studied the $X(3872)$-production
at PANDA, where two possible interpretations of $X(3872)$ have been
assumed. We have found that there will be a large number of events
for the process $p\bar p \to J/\psi \pi^+ \pi^-$
at the threshold where large fraction of events
will be produced from the decay of $X(3872)$.
By measuring the cross-section and its $s$-dependence near the threshold
one can distinguish the two interpretations.
For other possible interpretations
like a diquark anti-diquark bound state,
a hybrid charmonium state and a tetraquark state, the coupling with $p\bar p$
is so far unknown. Once the coupling is estimated, the production rate can be obtained
from our results here. If the coupling is not extremely small in comparison
with those given in Eq.(3,5), one may still expect that $X(3872)$ can be produced
with a not small event number.
Hence
the study of the $X(3872)$-production
at PANDA will provide important information about the
structure of $X(3872)$. We have also studied the production associated
with $\pi^0$. But the cross-section by considering the branching ratio
of $X(3872)$ may be too small to be measured.

\par
\vskip20pt
\noindent
{\bf\large  Acknowledgement:}
\par
We would like to
thank Prof. H.Q. Zheng, Dr. C. Meng and Dr. Y.J. Zhang for
helpful discussions. This work is supported by  National Nature
Science Foundation of P.R. China.
\par\vskip20pt


\end{document}